%% file: chaty_integraldublin2010.tex
\documentclass{PoS}

\usepackage{graphicx,txfonts,natbib}

\input{symbols}

\title{The INTEGRAL legacy on High Mass X-ray Binaries}

\ShortTitle{The INTEGRAL legacy on High Mass X-ray Binaries}

\author{\speaker{Sylvain Chaty}%
        \thanks{
We would like to thank the organisers for such a successfully organized and interesting workshop, in the historic Dublin castle! 
We are grateful to Ed van den Heuvel, and to all our close collaborators: 
P. Filliatre, P.-O. Lagage, Q.Z. Liu, I. Negueruela, L. Pellizza, F. Rahoui, J. Rodriguez, J. Tomsick, R. Walter and J.Z. Yan,
for many fruitful work and discussions on the study of obscured {\it INTEGRAL} sources.
This work was supported by the Centre National d'Etudes Spatiales (CNES), based on observations obtained with MINE --the Multi-wavelength INTEGRAL NEtwork--.
}\\
        Laboratoire AIM (UMR 7158 CEA/DSM-CNRS-Universit\'e Paris Diderot),
        Irfu/Service d'Astrophysique, CEA-Saclay, B\^at. 709,
        FR-91191 Gif-sur-Yvette Cedex, France \\
       E-mail: \email{sylvain.chaty@cea.fr}}

\author{Juan Antonio Zurita Heras\\
        François Arago Centre - Astroparticule \& Cosmologie (FACe/APC)
        Université Paris Diderot, CNRS/IN2P3, CEA/DSM, Observatoire de Paris
        13 rue Watt 75205 Paris Cedex 13, France \\
        E-mail: \email{zuritah@apc.univ-paris7.fr}}

\author{Arash Bodaghee\\
        Space Sciences Laboratory,
        University of California, Berkeley,
        7 Gauss Way, Berkeley, CA 94720, USA\\
        E-mail: \email{bodaghee@ssl.berkeley.edu}}

\abstract{Observations with the INTEGRAL satellite have quadrupled the population of supergiant High Mass X-ray Binaries (HMXBs), revealed a previously hidden population of obscured supergiant HMXBs, and allowed the discovery of huge and fast transient flares in supergiant HMXBs. 
Apart from these 3 observational facts, has INTEGRAL allowed us to better understand these supergiant HMXBs? Do we have now a better understanding of the 3 populations of HMXBs, and of their accretion process, separated in the so-called Corbet diagram? Do we better apprehend the accretion process in the supergiant HMXBs, and what makes the fast transient flares so special, in the context of the clumpy wind model, and of the formation of transient accretion disks? 
In summary, has the increased population of supergiant HMXBs allowed a better knowledge of these sources, compared to the ones that were already known before the launch of INTEGRAL? We will review all these observational facts, comparing to the current models, to objectively estimate what is the INTEGRAL legacy on High Mass X-ray Binaries.
}

\FullConference{8th INTEGRAL Workshop ``The Restless Gamma-ray Universe''\\
		 September 27-30 2010\\
		 Dublin Castle, Dublin, Ireland}

\begin{document}

\section{The $\gamma$-ray sky seen by the {\it INTEGRAL} satellite}

The {\it INTEGRAL} observatory is an ESA satellite launched on 17 October
2002 by a PROTON rocket on an excentric orbit. It is hosting 4
instruments: 2 $\gamma$-ray coded-mask telescopes --the imager IBIS
and the spectro-imager SPI, observing in the range 10 keV-10 MeV, with
a resolution of $12\amin$ and a field-of-view of $19\adeg$-- a
coded-mask telescope JEM-X (3-100 keV), and an optical telescope
(OMC).
%
%
The $\gamma$-ray sky seen by {\it INTEGRAL} is very rich, since 723 sources
have been detected by {\it INTEGRAL}, reported in the $4^{th}$ IBIS/ISGRI soft
$\gamma$-ray catalogue, spanning nearly 7 years of observations in the 17-100
keV domain \citep{bird:2010}\footnote{See an up-to-date list at {\em http://irfu.cea.fr/Sap/IGR-Sources/}, maintained by J. Rodriguez and A. Bodaghee}.  
%
Among these sources, there are 185 X-ray binaries (representing 26\% of the whole sample of sources detected by {\it INTEGRAL}, called ``IGRs'' in the following), 255 Active Galactic Nuclei (35\%), 35 Cataclysmic Variables (5\%), and $\sim 30$ sources of other type (4\%): 15 SuperNova Remnants, 4 Globular Clusters, 3 Soft $\gamma$-ray Repeaters, 2 $\gamma$-ray Burst, etc. 
215 objects still remain unidentified (30\%).

X-ray binaries are separated in 95 Low Mass X-ray Binaries (LMXBs) and 90 High Mass X-ray Binaries (HMXBs), each category representing $\sim 13$\% of IGRs.
Among identified HMXBs, there are 24 BeHMXBs (HMXBs hosting a Be companion star) and 19 sgHMXBs (HMXBs hosting a supergiant companion star), representing respectively 31\% and 24\% of HMXBs).
It is interesting to follow the evolution of the ratio between BeHMXBs
and sgHMXBs.  During the pre-{\it INTEGRAL} era, HMXBs were mostly BeHMXBs. For instance, in the catalogue of 130 HMXBs by
\citet{liu:2000}, there were 54 BeHMXBs and 7 sgHMXBs (respectively 42\% and 5\% of the total number of
HMXBs).  Then, the situation changed drastically with the first
HMXBs identified by {\it INTEGRAL}: in the catalogue of 114 HMXBs (+128 in the Magellanic Clouds) of \citet{liu:2006}, there
were 60\% of BeHMXBs and 32\% of sgHMXBs firmly identified. While the ratio of BeHMXBs/HMXBs increased by a factor of 1.5 only, the
sgHMXBs/HMXBs ratio increased by a factor of 6.




The most important result of {\it INTEGRAL} to date is the discovery of many new high energy sources -- concentrated in the Galactic plane, mainly towards tangential directions of Galactic arms, rich in star forming regions--, exhibiting common characteristics which previously had rarely been seen (see e.g. \citeauthor{chaty:2005a} \citeyear{chaty:2005a}). Many of them are HMXBs hosting a neutron star (NS) orbiting an OB companion, in most cases a supergiant star. Nearly all the {\it INTEGRAL} HMXBs for which both spin and orbital periods have been measured are located in the upper part of the Corbet diagramme (\citeauthor{corbet:1986} \citeyear{corbet:1986}, see Figure \ref{liufig}). They are wind accretors, typical of sgHMXBs, and X-ray pulsars exhibiting longer pulsation periods and higher absorption (by a factor $\sim4$) as compared to previously known HMXBs \citep{bodaghee:2007}. 
They divide into two classes: some are very obscured, exhibiting a huge intrinsic and local extinction, --the most extreme example being the highly absorbed source IGR~J16318-4848 \citep{filliatre:2004}--, and the others are HMXBs hosting a supergiant star and exhibiting fast and transient outbursts --an unusual characteristic among HMXBs--.  These are therefore called Supergiant Fast X-ray Transients (SFXTs, \citeauthor{negueruela:2006a} \citeyear{negueruela:2006a}),
with IGR~J17544-2619 being their archetype \citep{pellizza:2006}.

\section{Obscured HMXBs: IGR~J16318-4848, an extreme case}

  IGR~J16318-4848 was the first source discovered by IBIS/ISGRI on
  {\it INTEGRAL} on 29 January 2003 \citep{courvoisier:2003}, with a
  $2 \amin$ uncertainty.  {\it XMM-Newton} observations revealed a
  comptonised spectrum exhibiting an unusually high level of
  absorption: $\nh \sim 1.84 \times 10^{24} \cmmoinsdeux$
  \citep{matt:2003}.  The accurate localisation by {\it XMM-Newton}
  allowed \citet{filliatre:2004} to rapidly trigger ToO photometric and
  spectroscopic observations in optical/NIR, leading to the
  confirmation of the optical counterpart \citep{walter:2003} and to
  the discovery of the NIR one \citep{filliatre:2004}.  The extremely
  bright NIR source (B\,$>25.4\pm1$; I\,$=16.05\pm0.54$, J\,$= 10.33\pm 0.14$;
  H\,$=8.33\pm 0.10$ and Ks\,$=7.20 \pm 0.05 \mags$) exhibits an
  unusually strong intrinsic absorption in the optical ($A_v = 17.4
  \mags$), 100 times stronger than the interstellar absorption along
  the line of sight ($A_v = 11.4 \mags$), but still 100 times lower
  than the absorption in X-rays.  This led \citet{filliatre:2004} to
  suggest that the material absorbing in X-rays was concentrated
  around the compact object, while the material absorbing in
  optical/NIR was enshrouding the whole system.  

The NIR spectroscopy in the $0.95-2.5 \microns$ domain allowed \cite{filliatre:2004} to identify the nature of the companion star, by revealing an unusual spectrum, with many strong emission lines:
%
%
H, He${\rm I}$ (P-Cyg) lines --characteristic of dense/ionised wind at v\,$=400$\,km/s--,
He${\rm II}$ lines --signature of a highly excited region--,
$[$Fe${\rm II}]$ lines --reminiscent of shock heated matter--,
Fe${\rm II}$ lines --emanating from media of densities $>10^5-10^6$\,cm$^{-3}$-- and
Na${\rm I}$ lines --coming from cold/dense regions--.
%
%
All these lines originate from a highly complex, stratified
circumstellar environment of various densities and temperatures,
suggesting the presence of an envelope and strong stellar outflow
responsible for the absorption. Only luminous early-type stars such as
sgB[e] show such extreme environments, and
\citet{filliatre:2004} concluded that IGR~J16318-4848 was an unusual
HMXB hosting a sgB[e] with characteristic luminosity of
$10^6 \Lsol$ and mass of $30 \Msol$, located at a distance
between 1 and 6 kpc (see also \citeauthor{chaty:2005a} \citeyear{chaty:2005a}).
This source would therefore be the second HMXB hosting a sgB[e] star,
after CI Cam (see \citeauthor{clark:1999} \citeyear{clark:1999}). 

By combining optical, NIR and MIR observations, and fitting these observations with a model of sgB[e] companion star, \citet{rahoui:2008} showed that IGR~J16318-4848 was exhibiting a MIR excess, 
interpreted as due to the strong stellar outflow emanating from the sgB[e] companion star.  They found that this star had a temperature of $\Tstar=22200$\,K and radius $\Rstar = 20.4 \Rsol = 0.1$\,a.u., consistent with a supergiant star, and an extra component of temperature T $=1100$\,K and radius R\,$= 11.9\Rstar = 1$\,a.u., with A$_v = 17.6 \mags$.
%
%
Recent MIR spectroscopic observations with VISIR at the VLT showed
that the source was exhibiting strong emission lines of H, He, Ne, PAH, Si,
proving that the extra absorbing component was made of dust and
cold gas.
By assuming a typical orbital period of 10\,days and a mass of the
companion star of $20 \Msol$, we obtain an orbital separation of $50
\Rsol$, smaller than the extension of the extra component of dust/gas
($= 240 \Rsol$),
suggesting that this dense and absorbing circumstellar material envelope enshrouds the whole binary system, like 
a cocoon (see Figure \ref{figure:obscured-sfxt}, left panel).
We point out that this source exhibits such extreme
characteristics that it might not be fully representative of the other
obscured sources.

\section{Supergiant Fast X-ray Transients}


SFXTs constitute a new class of $\sim 12$ sources identified among the
recently discovered IGRs. They are HMXBs hosting NS orbiting around sgOB companion stars, exhibiting peculiar characteristics compared to ``classical'' HMXBs: rapid outbursts lasting only for hours, faint quiescent emission, and high energy spectra requiring a black hole (BH) or NS accretor. The flares rise in tens of minutes, last for $\sim$ 1 hour, their frequency is $\sim7$\,days, and their luminosity $L_x$ reaches $\sim 10^{36} \ergs$ at the outburst peak.
We can divide the SFXTs in two groups, according to the duration and
frequency of their outbursts, and their
$\frac{L_\mathrm{max}}{L_\mathrm{min}}$ ratio. Classical SFXTs
exhibit a very low quiescence $L_X$ and a high variability, while
intermediate SFXTs exhibit a higher $<L_X>$, a lower
$\frac{L_\mathrm{max}}{L_\mathrm{min}}$ and a smaller variability factor,
with longer flares.  SFXTs might appear like persistent sgHMXBs with
$<L_X>$ below the canonical value of $\sim 10^{36} \ergs$, and flares
superimposed.  But there might be some observational bias in these
general characteristics, therefore the distinction between SFXTs and
sgHMXBs is not well defined yet.
While the typical hard X-ray variability factor (the ratio between
deep quiescence and outburst flux) is less than 20 in
classical/absorbed systems, it is higher than 100 in SFXTs (some
sources can exhibit flares in a few minutes, like for instance
XTE\,J1739-302 \& IGR\,J17544-2619, see e.g. \citeauthor{pellizza:2006} \citeyear{pellizza:2006}). \\

To explain the emission of SFXTs in the context of sgHMXBs, 
\citet{negueruela:2008} and \citet{walter:2007} invoke the
existence of two zones around the supergiant star, of high and low
clump density respectively.  This would naturally explain the smooth
transition between sgHMXBs and SFXTs, and the existence of intermediate
systems; the main difference between classical sgHMXBs and SFXTs being
in this scenario the NS orbital radius.
Indeed, a basic model of porous wind predicts a substantial change in the
properties of the wind "seen by the NS" at a distance $r \sim 2
\Rstar$ (\citeauthor{negueruela:2008} \citeyear{negueruela:2008}), where we
stop seeing persistent X-ray sources. There are 2-regimes:
either the NS sees a large number of clumps, because it is
  embedded in a quasi-continuous wind;
or the number density of clumps is so small that the NS is effectively 
orbiting in an empty space.
The observed division between sgHMXBs (persistent sgHMXBs and SFXTs)
is therefore naturally explained by simple geometrical differences
in the orbital configurations.

\subsection{IGR\,J18483-0311: an intermediate SFXT?}

X-ray properties of this system were suggesting an SFXT nature
\citep{sguera:2007}, exhibiting however an unusual behaviour: its
outbursts last for a few days (to compare to hours for classical
SFXTs), and the ratio $L_{max}/L_{min}$ only reaches $\sim 10^3$ (meaning that its quiescence is at a higher level than the ratio $\sim 10^4$ for classical
SFXTs). Moreover, its orbital period $\Porb$=18.5d is low compared to
classical SFXTs (with large/eccentric orbits). Finally, its orbital
and spin periods ($P_\mathrm{spin}$=21.05s) located it ambiguously inbetween Be and
sgHMXBs in the Corbet Diagramme (see Figure \ref{liufig}).
\citet{rahoui:2008a} identified the companion star of this system
as a B0.5Ia supergiant, unambiguously showing that this system is an
SFXT.  Furthermore, they suggest that this system could be the first
firmly identified intermediate SFXT, characterised by short, eccentric
orbit (with an eccentricity $e$ between 0.4 and 0.6),
and long outbursts... An "intermediate"
SFXT nature would explain the unusual characteristics of this source among
"classical" SFXTs.

\subsection{What is the origin of ``misplaced'' sgHMXBs?}

\begin{figure}
\includegraphics[height=.3\textheight,angle=0]{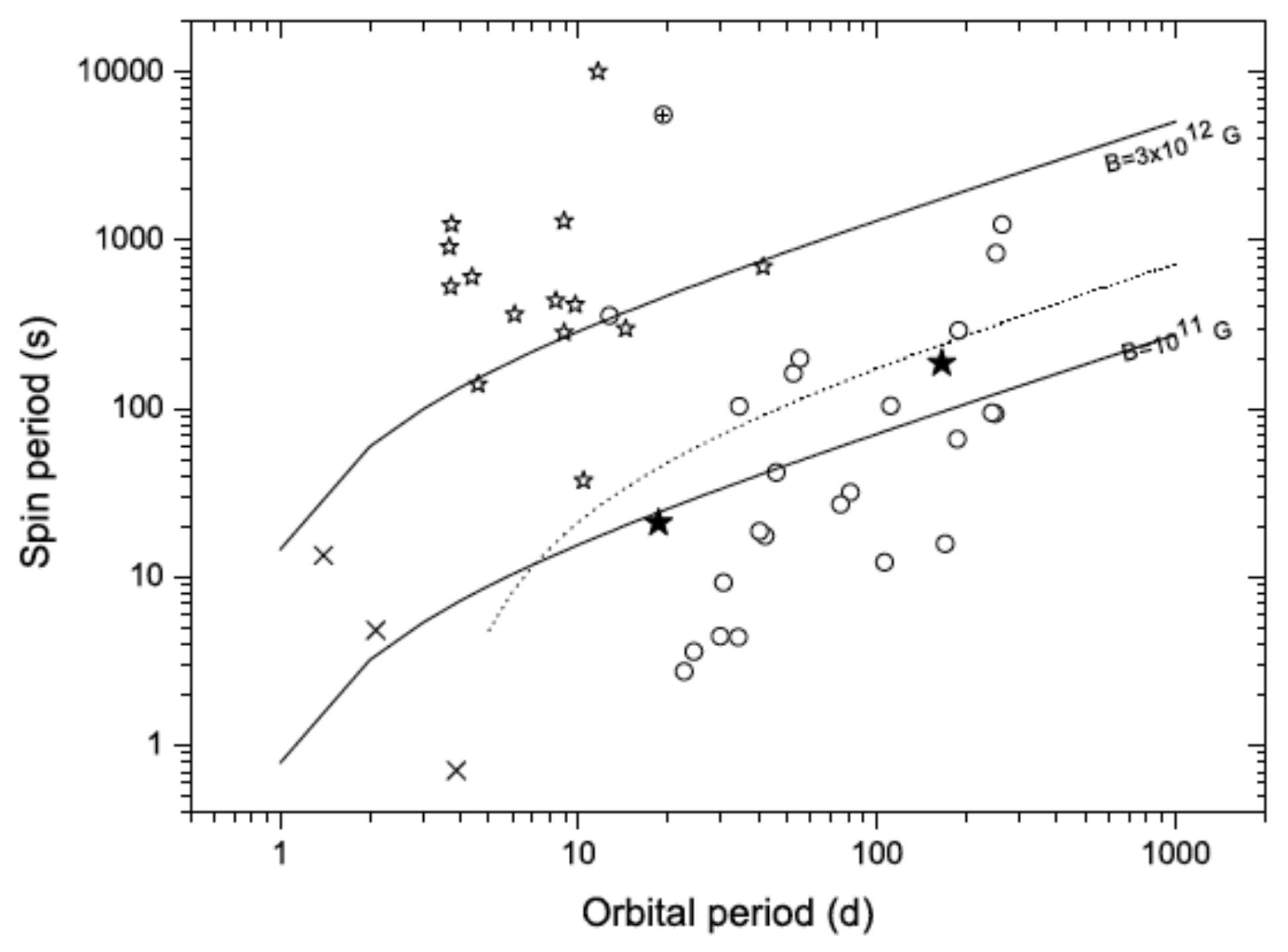}
\caption{\label{liufig}
Corbet diagram, adapted and taken from \citet{liu:2010}, showing the relation between orbital and spin periods. The open circles are for Be/X-ray binaries and the open stars for wind-fed sgHMXBs. The two ``misplaced'' SFXTs --IGR\,J11215-5952 and IGR\,J18483-0311-- are indicated by the filled stars. The solid lines stand for the theoretical equilibrium period for sgHMXBs, with magnetic field of $10^{11}$ and $3 \times 10^{12}$\,G, respectively. The dot line is the theoretical line with the parameters of B1 Ia supergiants (see \citeauthor{liu:2010} \citeyear{liu:2010} for more details).
}
\end{figure}

As noted by \citet{liu:2010}, there are two ``misplaced'' SFXTs in the Corbet diagram: IGR\,J11215-5952 (a neutron star orbiting a B1 Ia star, \citeauthor{negueruela:2005b} \citeyear{negueruela:2005b}) and IGR\,J18483-0311, described in the previous paragraph (see Figure \ref{liufig}). According to \citet{liu:2010}, these 2 SFXTs can not have evolved from normal Main Sequence O-type stars, since they are not at the equilibrium spin period of sgHMXBs (see e.g. \citeauthor{waters:1989} \citeyear{waters:1989}). They must therefore be the descendants of BeHMXBs (i.e. hosting O-type emission line stars), after the NS has reached the equilibrium spin period \citep{liu:2010}.

\section{The Grand Unification: different geometries, different scenarios}

In view of the results described above, 
there seems to be a continuous trend, from classical and/or absorbed
sgHMBs, to classical SFXTs. We outline in the following this trend.

\begin{figure}
\includegraphics[height=.275\textheight,angle=0]{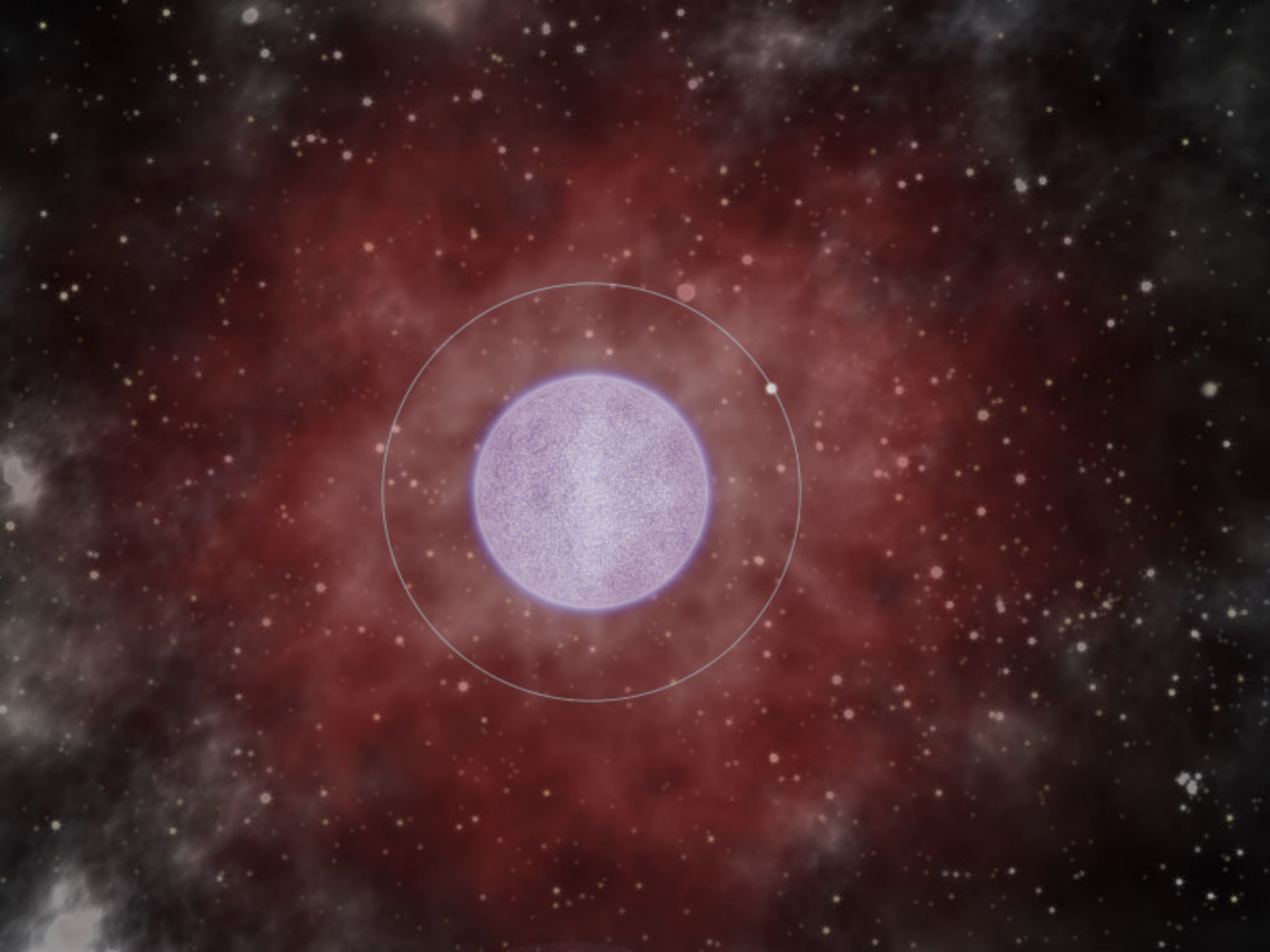}
\includegraphics[height=.275\textheight,angle=0]{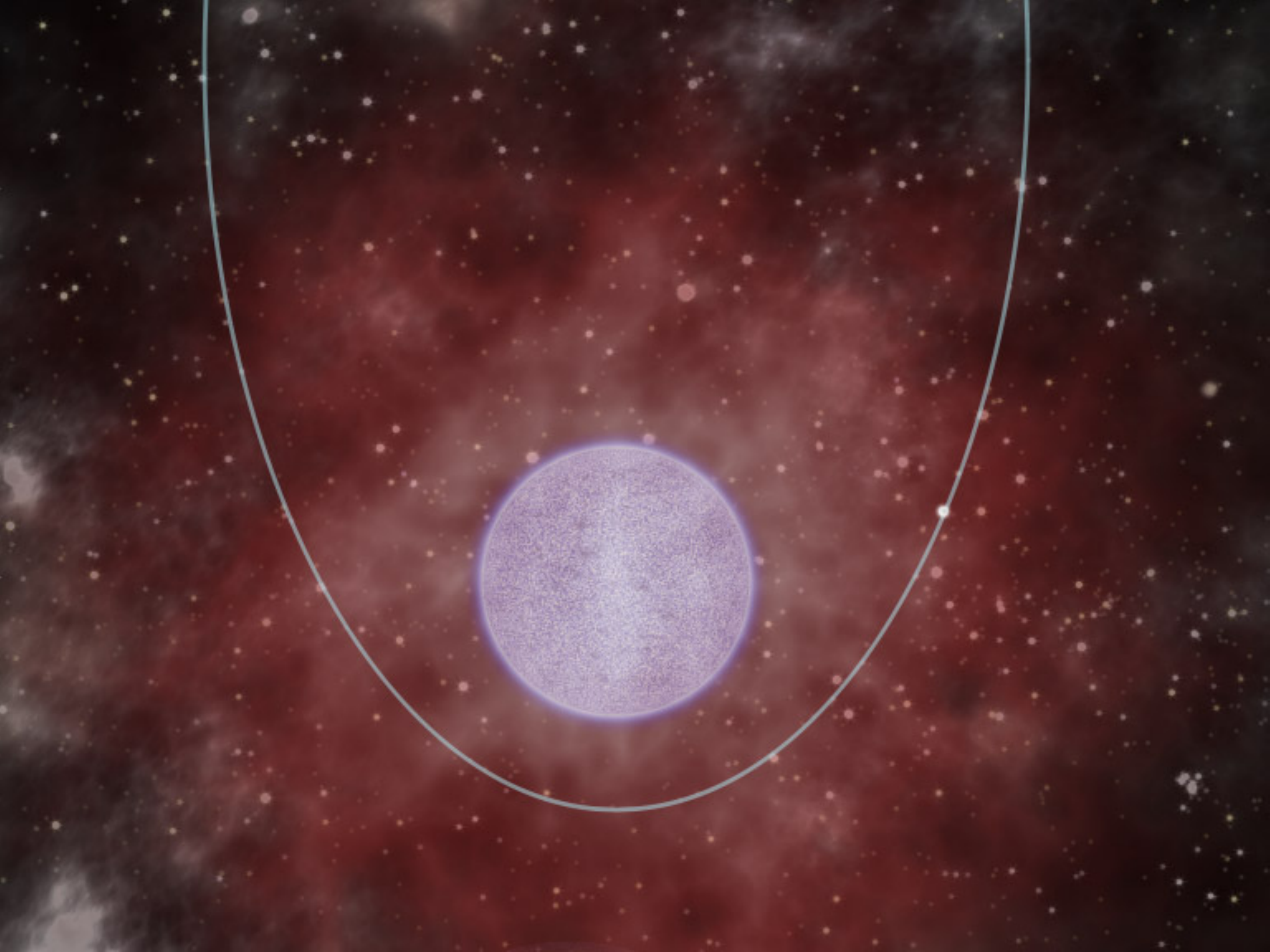}
\caption{\label{figure:obscured-sfxt}
  Scenarios illustrating two possible configurations of {\it INTEGRAL} 
  sources: a NS orbiting a supergiant
  star on a circular orbit (left image); and on an eccentric orbit
  (right image), accreting from the clumpy stellar wind of the
  supergiant.  The accretion of matter is persistent in the case of
  the obscured sources, as in the left image, where the compact object
  orbits inside the cocoon of dust enshrouding the whole system. On
  the other hand, the accretion is intermittent in the case of SFXTs,
  which might correspond to a compact object on an eccentric orbit, as
  in the right image.  A 3D animation of these sources is available on the
  website: {\em http://www.aim.univ-paris7.fr/CHATY/Research/hidden.html}
}
\end{figure}

\begin{description}

\item ["Classical" (or ``obscured'') sgHMXBs:]
  The absorbed (or obscured) sgHMXBs (like IGR\,J16318-4848) are classical sgHMXBs hosting NS constantly orbiting inside a cocoon made of dust and/or cold gas, probably created by the companion star itself. These systems therefore exhibit a persistent and luminous X-ray emission.  The cocoon, with an extension of $\sim 10 \Rstar = 1$\,a.u., is enshrouding the whole binary system. The NS, orbiting at a few stellar radii only from the star, has a short and circular orbit lying inside the zone of stellar wind high clump density ($R_{orb} \sim 2\Rstar$) (see Figure \ref{figure:obscured-sfxt}, left panel).

\item ["Intermediate" SFXT systems:] (such as IGR\,J18483-0311), the
  NS orbits on a short and circular/eccentric orbit, and it is only
  when the NS, orbiting inside the narrow transition zone, is close enough to the supergiant star that accretion takes place, and that X-ray emission arises, with possible periodic outbursts.

\item ["Classical" SFXTs:] (such as IGR\,J17544-2619), the NS orbits outside the high density zone, on a large and eccentric orbit around the supergiant star, and exhibits some  recurrent and short transient X-ray flares, while it comes close to the star, and accretes from clumps of matter coming from the wind of the supergiant.  Because it is passing through more diluted medium, the $\frac{Lmax}{Lmin}$ ratio is higher for "classical" SFXTs than for "intermediate" SFXTs (see Figure \ref{figure:obscured-sfxt}, right panel).

\end{description}

Although this scenario seems to describe quite well the characteristics
currently seen in sgHMXBs, we still need to identify the nature of
many more sgHMXBs to confirm it, and in particular the
orbital period and the dependance of the column density with the phase
of the binary system.

\section{Formation and evolution of sgHMXBs, link with population synthesis models}

sgHMXBs revealed by {\it INTEGRAL} will allow us to better constrain and understand the formation and evolution of X-ray binary systems, by comparing them to numerical study of LMXB/HMXB population synthesis models. For instance, these new systems might represent a precursor stage of what is known as the "Common Envelope phase" in the evolution of LMXB/HMXB systems, when the orbit has shrunk so much that the neutron star begins to orbit inside the envelope of the supergiant star. In addition, many parameters do influence the various evolutions, from one system to another: differences in mass, size, orbital period, ages, rotation, magnetic field, accretion type, stellar endpoints, etc... Moreover, stellar and circumstellar properties also influence the evolution of high-energy binary systems, made of two massive components likely born in rich star forming regions.

In the very nice review on the formation and evolution of relativistic binaries written by \citet{vandenHeuvel:2009}, the evolution of these supergiant INTEGRAL sources is mentioned. We can have an idea of the formation of these systems, since orbital periods of later evolutionary phases are linearly dependent of initial orbital periods. It is therefore possible to derive that the initial orbital periods of currently very wide O-supergiant INTEGRAL binaries could be as long as 100 days \cite{vandenHeuvel:2009}.
The systems having long orbital periods are expected to survive the
Common Envelope phase. They may then either end as a close eccentric double neutron star, or in some cases, as a black hole-neutron star binary
 (van den Heuvel, priv. comm.).

No such system is known yet, however some of these systems might harbour a black hole as the compact object. Of course neutron stars are easier to detect through X-ray pulsations, but, as Carl Sagan already pointed it out, {\it absence of evidence is not evidence of absence}... We should look for black holes orbiting around supergiant companion stars in wind-accreting HMXBs, however this is only feasible through observational methods involving detection of extremely faint radial velocity displacement due to the high mass of the companion star, or through extremely accurate radio measurements that will be available in the future. On the other hand, massive stars lose so much matter during their evolution that they might always finish as neutron stars (see e.g. \citeauthor{maeder:2008} \citeyear{maeder:2008}). If this is the case, then such systems hosting black holes might not form at all. 

Finally, these sources are also useful to look for massive stellar
"progenitors", for instance giving birth to coalescence of compact
objects, through NS/NS or NS/BH collisions. They would then become
prime candidate for gravitational wave emitters, or even to short/hard
$\gamma$-ray bursts.

\section{Conclusions and perspectives...}

The {\it INTEGRAL} satellite has quadrupled the total number of Galactic
sgHMXBs, constituted of a NS orbiting around a supergiant
star. Most of these new sources are slow and absorbed X-ray pulsars,
exhibiting a large $\nh$ and long $P_\mathrm{spin}$ ($\sim$1ks).  The
influence of the local absorbing matter on periodic modulations is
different for sgHMXBs or BeHMXBs, segregated in
different parts of $\nh$-$\Porb$ or $\nh$-$P_\mathrm{spin}$.
%
{\it INTEGRAL} revealed 2 new types of sources.  First, the SFXTs, exhibiting short and strong X-ray
flares, with a peak flux of 1 Crab during 1--100s, every $\sim 100$\,days.
These flares can be explained by accretion through
clumpy winds.  Second, the obscured HMXBs are persistent X-ray sources
composed of supergiant stellar companions exhibiting a strong intrinsic absorption and long $P_\mathrm{spin}$.  The NS is
deeply embedded in the dense stellar wind, forming a dust cocoon
enshrouding the whole binary system.

These results show the existence in our Galaxy of a dominant
population of a previously rare class of high-energy binary systems:
sgHMXBs, some of them exhibiting a high intrinsic absorption
(\citeauthor{chaty:2008} \citeyear{chaty:2008};
\citeauthor{rahoui:2008} \citeyear{rahoui:2008}). Studying
this population will provide a
better understanding of the formation and evolution of short-living
HMXBs.  Furthermore, stellar population models now have to
take these objects into account, to assess a realistic number of
high-energy binary systems in our Galaxy. 





\input{chaty_integraldublin2010.bbl}

\end{document}

%% file: symbols.tex
\def\cmmoinsdeux{\mbox{ cm}^{-2}}

\def\microns{\mbox{ } \mu \mbox{m}}

\def\Msol{\mbox{ }M_{\odot}}
\def\Rsol{\mbox{ }R_{\odot}}
\def\Lsol{\mbox{ }L_{\odot}}

\def\Rstar{\mbox{ }R_{\star}}
\def\Tstar{\mbox{ }T_{\star}}

\def\Porb{P_{orb}}

\def\mags{\mbox{ magnitudes}}

\def\ergs{\mbox{ erg\,s}^{-1}}

\def\hour{^{h}}

\def\adeg{^{\circ}}

\def\amin{^\prime}

\def\nh{N_{\rm H}}

\def\ltsima{\; \buildrel < \over \sim \;}
\def\simlt{\lower.5ex\hbox{\ltsima}}            
\def\gtsima{\; \buildrel > \over \sim \;}
\def\simgt{\lower.5ex\hbox{\gtsima}}            


%% file: chaty_integraldublin2010.bbl
\begin{thebibliography}{24}
\expandafter\ifx\csname natexlab\endcsname\relax\def\natexlab#1{#1}\fi

\bibitem[{{Bird} {et~al.}(2010){Bird}, {Bazzano}, {Bassani}, {Capitanio},
  {Fiocchi}, {Hill}, {Malizia}, {McBride}, {Scaringi}, {Sguera}, {Stephen},
  {Ubertini}, {Dean}, {Lebrun}, {Terrier}, {Renaud}, {Mattana}, {G{\"o}tz},
  {Rodriguez}, {Belanger}, {Walter}, \& {Winkler}}]{bird:2010}
{Bird}, A.~J., {Bazzano}, A., {Bassani}, L., {et~al.} 2010, ApJSS, 186, 1

\bibitem[{{Bodaghee} {et~al.}(2007){Bodaghee}, {Courvoisier}, {Rodriguez},
  {Beckmann}, {Produit}, {Hannikainen}, {Kuulkers}, {Willis}, \&
  {Wendt}}]{bodaghee:2007}
{Bodaghee}, A., {Courvoisier}, T.~J.-L., {Rodriguez}, J., {et~al.} 2007, A\&A,
  467, 585

\bibitem[{{Chaty} \& {Filliatre}(2005)}]{chaty:2005a}
{Chaty}, S. \& {Filliatre}, P. 2005, A\&ASS, 297, 235

\bibitem[{{Chaty} {et~al.}(2008){Chaty}, {Rahoui}, {Foellmi}, {Rodriguez},
  {Tomsick}, \& {Walter}}]{chaty:2008}
{Chaty}, S., {Rahoui}, F., {Foellmi}, C., {et~al.} 2008, A\&A, 484, 783

\bibitem[{{Clark} {et~al.}(1999){Clark}, {Steele}, {Fender}, \&
  {Coe}}]{clark:1999}
{Clark}, J.~S., {Steele}, I.~A., {Fender}, R.~P., \& {Coe}, M.~J. 1999, A\&A,
  348, 888

\bibitem[{{Corbet}(1986)}]{corbet:1986}
{Corbet}, R.~H.~D. 1986, MNRAS, 220, 1047

\bibitem[{{Courvoisier} {et~al.}(2003){Courvoisier}, {Walter}, {Rodriguez},
  {Bouchet}, \& {Lutovinov}}]{courvoisier:2003}
{Courvoisier}, T.~J.-L., {Walter}, R., {Rodriguez}, J., {Bouchet}, L., \&
  {Lutovinov}, A.~A. 2003, IAU Circ., 8063, 3

\bibitem[{{Filliatre} \& {Chaty}(2004)}]{filliatre:2004}
{Filliatre}, P. \& {Chaty}, S. 2004, ApJ, 616, 469

\bibitem[{{Liu} {et~al.}(2010){Liu}, {Chaty}, \& {Yan}}]{liu:2010}
{Liu}, Q.~Z., {Chaty}, S., \& {Yan}, J. 2010, ApJ subm.

\bibitem[{{Liu} {et~al.}(2000){Liu}, {van Paradijs}, \& {van den
  Heuvel}}]{liu:2000}
{Liu}, Q.~Z., {van Paradijs}, J., \& {van den Heuvel}, E.~P.~J. 2000, A\&ASS,
  147, 25

\bibitem[{{Liu} {et~al.}(2006){Liu}, {van Paradijs}, \& {van den
  Heuvel}}]{liu:2006}
{Liu}, Q.~Z., {van Paradijs}, J., \& {van den Heuvel}, E.~P.~J. 2006, A\&A,
  455, 1165

\bibitem[{{Maeder} \& {Meynet}(2008)}]{maeder:2008}
{Maeder}, A. \& {Meynet}, G. 2008, in Revista Mexicana de Astronomia y
  Astrofisica Conference Series, Vol.~33, Revista Mexicana de Astronomia y
  Astrofisica Conference Series, 38--43

\bibitem[{{Matt} \& {Guainazzi}(2003)}]{matt:2003}
{Matt}, G. \& {Guainazzi}, M. 2003, MNRAS, 341, L13

\bibitem[{{Negueruela} {et~al.}(2005){Negueruela}, {Smith}, \&
  {Chaty}}]{negueruela:2005b}
{Negueruela}, I., {Smith}, D.~M., \& {Chaty}, S. 2005, The Astronomer's
  Telegram, 470, 1

\bibitem[{{Negueruela} {et~al.}(2006){Negueruela}, {Smith}, {Reig}, {Chaty}, \&
  {Torrej{\'o}n}}]{negueruela:2006a}
{Negueruela}, I., {Smith}, D.~M., {Reig}, P., {Chaty}, S., \& {Torrej{\'o}n},
  J.~M. 2006, in ESA Special Publication, Vol. 604, ESA Special Publication,
  ed. A.~{Wilson}, 165--170

\bibitem[{{Negueruela} {et~al.}(2008){Negueruela}, {Torrejon}, {Reig}, {Ribo},
  \& {Smith}}]{negueruela:2008}
{Negueruela}, I., {Torrejon}, J.~M., {Reig}, P., {Ribo}, M., \& {Smith}, D.~M.
  2008, ArXiv e-prints, 801

\bibitem[{{Pellizza} {et~al.}(2006){Pellizza}, {Chaty}, \&
  {Negueruela}}]{pellizza:2006}
{Pellizza}, L.~J., {Chaty}, S., \& {Negueruela}, I. 2006, A\&A, 455, 653

\bibitem[{{Rahoui} \& {Chaty}(2008)}]{rahoui:2008a}
{Rahoui}, F. \& {Chaty}, S. 2008, ArXiv e-prints

\bibitem[{{Rahoui} {et~al.}(2008){Rahoui}, {Chaty}, {Lagage}, \&
  {Pantin}}]{rahoui:2008}
{Rahoui}, F., {Chaty}, S., {Lagage}, P.-O., \& {Pantin}, E. 2008, A\&A, 484,
  801

\bibitem[{{Sguera} {et~al.}(2007){Sguera}, {Hill}, {Bird}, {Dean}, {Bazzano},
  {Ubertini}, {Masetti}, {Landi}, {Malizia}, {Clark}, \&
  {Molina}}]{sguera:2007}
{Sguera}, V., {Hill}, A.~B., {Bird}, A.~J., {et~al.} 2007, A\&A, 467, 249

\bibitem[{{van den Heuvel}(2009)}]{vandenHeuvel:2009}
{van den Heuvel}, E.~P.~J. 2009, in Astrophysics and Space Science Library,
  Vol. 359, Astrophysics and Space Science Library, ed. {M.~Colpi, P.~Casella,
  V.~Gorini, U.~Moschella, \& A.~Possenti }, 125--+

\bibitem[{{Walter} {et~al.}(2003){Walter}, {Rodriguez}, {Foschini}, {de Plaa},
  {Corbel}, {Courvoisier}, {den Hartog}, {Lebrun}, {Parmar}, {Tomsick}, \&
  {Ubertini}}]{walter:2003}
{Walter}, R., {Rodriguez}, J., {Foschini}, L., {et~al.} 2003, A\&A, 411, L427

\bibitem[{{Walter} \& {Zurita Heras}(2007)}]{walter:2007}
{Walter}, R. \& {Zurita Heras}, J. 2007, A\&A, 476, 335

\bibitem[{{Waters} \& {van Kerkwijk}(1989)}]{waters:1989}
{Waters}, L.~B.~F.~M. \& {van Kerkwijk}, M.~H. 1989, A\&A, 223, 196

\end{thebibliography}
